\documentclass[11pt]{article}
\usepackage[letterpaper,hmargin=1in,vmargin=1.25in]{geometry}

\author{Hao Chen\\
Department of Computing and \\
Information Technology\\
School of Information Science\\
and Engineering\\
Fudan University\\
Shanghai,200433\\
People's Republic of China}
\title{\bf Linear Secret Sharing  from \\
Algebraic-Geometric Codes}
\date{December, 2005}

\begin{document}

\maketitle
\begin{abstract}
It is well-known that the linear secret-sharing scheme (LSSS) can be
constructed from  linear error-correcting codes (Brickell [1], R.J.
McEliece and D.V.Sarwate [2],Cramer, el.,[3]). The theory of linear
codes from algebraic-geometric curves (algebraic-geometric (AG)
codes or geometric Goppa code) has been well-developed since the
work of V.Goppa and Tsfasman, Vladut, and Zink( see [17], [18] and
[19]). In this paper the linear secret-sharing scheme from
algebraic-geometric codes, which are non-threshold schemes for
curves of genus greater than 0, are presented . We analysis the
minimal access structure, $d_{min}$ and $d_{cheat}$([8]), (strongly)
multiplicativity and the applications in verifiable secret-sharing
(VSS) scheme and secure multi-party computation (MPC) of this
construction([3] and [10-11]). Our construction also offers many
examples of the self-dually $GF(q)$-representable matroids and many
examples of new ideal linear secret-sharing schemes addressing to
the problem of the characterization of the access structures for
ideal secret-sharing schemes([3] and [9]). The access structures of
the linear secret-sharing schemes from the codes on elliptic curves
are given explicitly. From the work in this paper we can see that
the algebraic-geometric structure of the underlying algebraic curves
is an important resource for secret-sharing,
matroid theory, verifiable secret-sharing and secure multi-party computation.   \\


\end{abstract}

{\bf I. Introduction and Preliminaries}\\

In a secret-sharing scheme among the set of participants ${\bf
P}=\{P_1,...,P_n\}$, a dealer $P_0$, not in ${\bf P}$, has a secret,
the dealer distributes the secret among ${\bf P}$, that is gives
each participant a share of secret, in such a way
 that only the qualified subsets of  ${\bf P}$ can reconstruct the secret from their shares.
 The access structure , $\Gamma \subset 2^{{\bf P}}$, of a secret-sharing scheme is the defined to be the family of
 the qualified subsets of ${\bf P}$. The minimum accesss structure $min
 \Gamma \subset 2^{{\bf P}}$ is defined to the be the set of minimum elements in
 $\Gamma$(here we use the natural order relation $ S_1 <S_2$ if and
 only if $ S_1 \subset S_2$ on $2^{{\bf P}}$). The family of all subsets of ${\bf P}$
 which are not qualified is called the adversary structure( see [3] and {10]). We call a secret-sharing scheme a $(k,n)$-threshold scheme if the access
 structure
 consists of the subset of at least  $k$ elements in the set ${\bf P}$, where the number of elements in the set ${\bf P}$
 is
 exactly $n$, that is, among the $n$ members any subset of $k$ or more than $k$ members can reconstruct the secret.
The first secrets-sharing scheme was given independently by Blakley
[4] and Shamir [5] in 1979, actually they gave
 threshold secret-sharing scheme. We call a secret-sharing scheme perfect if the the unqualified
subsets of members to reconstruct the secret have no information of
the secret. The existence of secret-sharing schemes with arbitrary given access structures was proved in [6] and [7].\\

For a secret-sharing scheme, let ${\bf V}$ be the set of all
possible shares $(v_1,...,v_n)$ (Here $v_i$ is the share of the
participant $P_i$ for $i=1,...,n$). Then ${\bf V}$ is a
error-correcting code(not necessarily linear), let $d_{min}$ be the
minimum Hamming distance of this error-correcting code ${\bf V}$.
From the error-correcting capability, it is clear that the cheaters
can be identified from any share(presented by the
participants)$(v_1,...,v_n)$ if there are at most $[(d_{min}-1]/2]$
cheaters. In [2] McEliece and Sarwate proved that $d_{min}=n-k+1$
for Shamir's $(k,n)$-threshold scheme. K.Okada and K.Kurosawa
introduced anther parameter $d_{cheat}$ for general secret-sharing
scheme, as the the number such that the correct secret value $s$ can
be recovered if there are at most $[(d_{cheat}-1)/2]$ cheaters
([8]). It is clear $d_{min} \leq d_{cheat}$, it is proved in [8]
that $d_{cheat} = n-max_{B \in (2^{{\bf P}}-\Gamma})|B|$, where
$|B|$ is the number of
the elements in the set $B$. \\

Let $K$ be a finite field. A $K$-linear secret sharing scheme (LSSS)
on the set of participants ${\bf P}=\{P_1,...,P_n\}$ is defined as a
sequences of surjective linear mappings $\{T_0,T_1,...,T_n\}$, where
$T_i: E \longrightarrow E_i$, $E$ and $E_i$ are finite dimensional
spaces over $K$($E_0=K$). For any $x \in E$, $\{T_1(x),...,T_n(x)\}$
are the shares of of the secret value $k=T_0(x)$. The complexity of
the $K$-LSSS is defined as  $\lambda(\Gamma)=\Sigma_{i=1}^n
dim_K(E_i)$, when the complexity is $n$, this LSSS is called {\em
ideal}. One of the main open problem in secrete sharing is the
characterization of
the access structures of ideal secret sharing schemes (see [3] and [9]).\\

For an access structure $\Gamma$, $\lambda_{K}(\Gamma)$ is defined
to be the minimum of all the values  $\Sigma_{i=1}^n dim_{K}(E_i)$
for $K$-linear
 secret sharing schemes with access structure $\Gamma$(see [12-13]).A LSSS is called
 {\em multiplicative, $K$-MLSSS} if every participant $i \in P$ can compute,
 form his shares $k_i,k_i'$ of two
 shared secrets $k,k'\in K$, a value $c_i \in K$ such that the product $kk'$ is a
 linear combination of all the values $c_1,...,c_n$. It is called
 strongly multiplicative if for any subset $A$ such that ${\bf P} -A$ is not
 qualified, the product $kk'$ can be computed using only values from the
 participants in $A$. $\mu_{K}(\Gamma)$ is defined to
 be the minimum of all the values of $\Sigma_{i=1}^n dim_{K}(E_i)$ for multiplicative
 $K$-linear secret sharing schemes with access structure $\Gamma$.
For an access structure $\Gamma$ on ${\bf P}$, it is said that
$\Gamma$ is $Q_2$ if $A\bigcup B\neq {\bf P}$ for any $A,B \in
\Gamma$, $\Gamma$ is $Q_3$ if $A\bigcup B \bigcup C\neq {\bf P}$ for
any $A,B,C \in \Gamma$. One of the key result in [10] is a method to
construct, from any LSSS with $Q_2$ access structure $\Gamma$, a
multiplicative LSSS with the same access structure and double
complexity, that is $\mu(\Gamma) \leq 2\lambda (\Gamma)$. $K$-MLSSS
and $Q_2$ ,$Q_3$
access structure are closely related to secure multiparty computations (see [3],[10] and [11]).\\

The approach of secret-sharing based on error-correcting codes was
studied in [1],[2],[3],[12-15]. It is found that actually Shamir's
$(k,n)$-threshold scheme is just the secret-sharing scheme based on
the famous Reed-Solomon (RS) code. The error-correcting code based
 secret-sharing scheme is defined as follow. Here we suppose ${\bf C}$ is
 a error-correcting codes over the finite
field $GF(q)$ (where $q$ is a prime power) with code length $n+1$
and dimension $k$, i.e., ${\bf C}$ is a $k$ dimension subspace of
$GF(q)^{n+1}$  The Hamming distance $d({\bf C})$ of this
error-correcting code ${\bf C}$ is defined as follows.\\

$$
\begin{array}{ccccccc}
d({\bf C})=min\{wt(v): v \in {\bf C}\}\\
wt(v)=|\{i:v=(v_0,v_1,...,v_n),v_i \neq 0\}|
\end{array}
$$
,where $wt(v)$ is called the Hamming weight of $v$. Let
$G=(g_{ij})_{1 \leq i \leq k ,0 \leq j \leq n}$ be the generator
matrix of ${\bf C}$, i.e., $G$ is a $k \times (n+1)$ matrix in which
$k$ rows of $G$ is a base of the $k$ dimension subspace ${\bf C}$ of
$GF(q)^{n+1}$. Suppose $s$ is a given secret value of the dealer
$P_0$ and the secret is shared among ${\bf P}=\{P_1,...,P_n\}$, the
set of $n$ participants .
 Let ${\bf g_1}=(g_{11},...,g_{k1})^T$ be the 1st column of $G$. Chosen a
 random ${\bf u}=(u_1,...,u_k) \in GF(q)^k$ such that $s={\bf u}^{\tau} {\bf g_0}=\Sigma u_ig_{i0}$.
 The codeword ${\bf c}=(c_0,...,c_N)={\bf u}G$, it is clear that $c_0=s$ is the secret, then the dealer $P_0$
 gives the $i-th$ participant $P_i$ the $c_i$ as the share of $P_i$ for $i=1,...,n$. In this
 secret-sharing scheme the error-correcting code ${\bf C}$ is assumed to be known to every participant and the dealer.
 For a secret sharing scheme form error-correcting codes, suppose that $T_i:GF(q)^k \longrightarrow GF(q)$
  is defined as $T_i({\bf x})= {\bf x}^{\tau} {\bf g_i}$, where $i=0,...,n$ and ${\bf g_i}$ is the $i$-th column of the
  generator matrix of the code ${\bf C}$. In this form we see that the secret sharing scheme is an ideal  $GF(q)$-LSSS. \\

We refer the following Lemma to [12-15].\\

 {\em {\bf Lemma 1 }(see [12-15]).
  Suppose the dual of ${\bf C}$,
 ${\bf C^{\perp}}=\{v=(v_0,..,v_n): Gv=0\}$ has no codeword of Hamming
 weight 1. In the above secret-sharing scheme based on the error-correcting code
 ${\bf C}$, $(P_{i_1},...,P_{i_m})$ can reconstruct the secret if and only
 if there is a codeword $v=(1,0,...,v_{i_1},...,v_{i_m},...0)$ in ${\bf
 C^{\perp}}$
 such that $v_{i_j} \neq 0$ for at least one $j$, where $1 \leq j \leq m $.}\\

 The secret reconstruction is as follows, since $Gv=0$, ${\bf
 g_1}=-\Sigma_{j=1}^m v_{i_j} {\bf g_{i_j}}$, where ${\bf g_h}$ is
 the $h-th$ column of $G$ for $h=1,...,N$. Then $s=c_0={\bf u}{\bf
 g_1}=-{\bf u}\Sigma_{j=1}^m  {\bf g_{i_j}}=-\Sigma_{j=1}^m
 v_{i_j}c_{i_j}$.\\

For the definition of matroid and the matroid on the set ${\bf
P}=\{P_1,...,P_n\}$ from a linear $[n,k,d]$ code, we refer to [16],
it is well-known that the circuits (minimal dependent set) on the
matroid from a linear code is in one-to-one correspondence of the
codewords with minimum Hamming weight $d$. Thus a subset $A$ of
${\bf P}=\{P_1,...,P_n\}$
 is a minimal qualified subset of the LSSS from linear code ${\bf C}$ if and only
 if $\{0\} \bigcup A$ is a circuit in the matroid from ${\bf C}$ (see [3]).   \\

We need recall some basic facts about algebraic-geometric codes( see
[17],[18] and [19]). Let ${\bf X}$ be an absolutely irreducible,
projective and smooth curve defined over $GF(q)$ with genus $g$,
${\bf D}=\{P_0,...P_n\}$ be a set of $GF(q)$-rational points of
${\bf X}$ and ${\bf G}$ be a $GF(q)$-rational divisor satisfying
$supp({\bf G})\bigcap {\bf D}=\empty$, $2g-2 < deg({\bf G}) <n+1$.
Let $L(G)=\{f: (f)+G \geq 0 \}$ is the linear space (over $GF(q)$)
of all rational functions with its divisor not smaller than $-G$ and
$\Omega(B)=\{\omega: (\omega) \geq B\}$ be the linear space of all
differentials with their divisors not smaller than $B$. Then the
functional AG(algebraic-geometric )code ${\bf C_L(D,G)} \in
GF(q)^{n+1}$ and residual AG(algebraic-geometric) code ${\bf
C_{\Omega}(D,G)} \in GF(q)^{n+1}$ are defined. ${\bf C_L(D,G)}$ is a
$[n+1,k=deg({\bf G})-g+1, d \geq n+1-deg({\bf G})]$ code over
$GF(q)$ and ${\bf C_{\Omega}(D,G)}$ is a $[n+1,k=n-deg({\bf G})+g,d
\geq deg({\bf G})-2g+2]$ code over $GF(q)$. We know that the
functional code is just the evaluations of functions in $L(G)$ at
the set ${\bf D}$ and the residual code is just the residues of
differentials in $\Omega(G-D)$ at the set ${\bf D}$ (see [17-19]).\\

We also know that ${\bf C_L(D,G)}$ and ${\bf C_{\Omega}(D,G)}$ are
dual codes. It is known that for a differential $\eta$ that has
poles at $P_1,...P_n$ with residue 1 (there always exists such a
$\eta$, see[18]) we have ${\bf C_{\Omega}(D,G)}={\bf
C_L(D,D-G+(\eta))}$, the function $f$ corresponds to the
differential $f\eta$. This means that  functional codes and residue
code are essentially the same. It is clear that if there exist a
differential $\eta$ such that ${\bf G}={\bf D}-{\bf G}+(\eta)$, then
${\bf C_L(P,G)}={\bf C_{\Omega}(P,G)}={\bf C_L(P,P-G+(\eta))}$ is a
self-dual code over $GF(q)$, in many cases the matroids from
AG-codes can be computed explicitly from the algebraic-geometric
structure of the underlying curves, thus we have many interesting
examples of self-dually $GF(q)$-representable matroids (see section
VI below)from this construction. For many examples of AG codes,
including these self-dual AG-codes,
we refer to [17],[18] and [19].\\

{\bf II Main Results}\\

Let ${\bf X}$ be an absolutely irreducible, projective and smooth
curve defined over $GF(q)$ with genus $g$, ${\bf D}=\{P_0,...P_n\}$
be a set of $GF(q)$-rational points of ${\bf X}$ and ${\bf G}$ be a
$GF(q)$-rational divisor with degree $m$ satisfying $ supp({\bf G})
\bigcap {\bf D}=\empty$, $2g-2 < m <n+1$. We can have a LSSS on the
$n$ participants ${\bf P}=\{P_1,...,P_n\}$ from the linear code
${\bf C_{\Omega}(D,G)}$, thus we know that the reconstruction of the
secret is based from its dual code ${\bf C_L(D,G)}$. For the curve
with genus 0 over $GF(q)$, we have exactly the same LSSS as Shamir's
$(k,n)$-threshold scheme, since the AG-codes over the curve of genus
0 is just the RS codes
(see [17-19]).\\

The following are the main results of this paper.\\

{\em {\bf Theorem 1.} The LSSS over $GF(q)$ from the code ${\bf
C_{\Omega}(D,G)}$ has the following properties.\\
1) This LSSS is ideal;\\
2) Any subset $A \subseteq P$ satisfying $|A| <n-m$ is not qualified
subset, any subset $A \subseteq P$ satisfying $|A| \geq n-m+2g$ is
qualified.}\\

{\em {\bf Proposition 1.} Let ${\bf X}, {\bf D},{\bf P}$ and ${\bf
G}$ as above. Suppose the genus $g$ of ${\bf X}$ is not $0$, $n> 3$
and the minimum (Hamming) distance $d({\bf C_L(D,G)})$ is exactly
$n+1-deg({\bf G})>2 $. Then the LSSS from the residue code ${\bf
C_{\Omega}(D,G)}$ is not a threshold secret-sharing scheme.}\\

{\em {\bf Proposition 2.} For the LSSS over $GF(q)$ from the code
${\bf C_L(D,G)}$ we have $n-m \leq d_{min} \leq d_{cheat} \leq
n-m+2g$.}\\

{\em {\bf Theorem 2.} The LSSS over $GF(q)$ from the code ${\bf
C_{\Omega}(D,G)}$ has the following properties.\\
1) This LSSS is multiplicative if $m \geq \frac{n}{2}+2g$;\\
2) This LSSS is strongly multiplicative if $m \geq \frac{2n}{3}+2g$ .}\\

Let ${\bf X},{\bf D},{\bf P}$ and ${\bf G}$ as above, $\emph{A}$ is
the adversary structure of the LSSS from the residue code ${\bf
C_{\Omega}(D,G)}$. Then we have the following result
(for the definitions in the following result we refer to [10] and [11]).\\

{\em {\bf Theorem 3.} For the finite field $GF(q)$ and the adversary
structure $\emph{A}$ as above.\\
1) If $m \geq \frac{2n}{3}+2g$, then there exists a polynomial
complexity error-free VSS (Verifiable Secret-Sharing , over
$GF(q)$)protocol in the information-theoretic scenario, secure
against any active and
adaptive $\emph{A}$-adversary.\\
2) If $m \geq \frac{2n}{3}+2g$, for any arithmetic circuit $U$ over
$GF(q)$, then there exists a polynomial complexity error-free MPC
(Multi-party Computation)protocol computing $U$ in the
information-theoretic scenario, secure against any adaptive and
active $\emph{A}$-adversary.\\
3) If $m \geq \frac{n}{2}+2g$, for any arithmetic circuit $U$ over
$GF(q)$, then there exists a polynomial complexity error-free MPC
(Multi-party Computation)protocol computing $U$ in the
information-theoretic scenario, secure against any adaptive and
passive $\emph{A}$-adversary.}\\

{\bf Proof of Theorem 1.} 1) is clear from the construction. If $A$
is a qualified subset of ${\bf P}$, then there exists a codeword in
${\bf C_L(D,G)}$ such that this codeword is not zero at $P_0$ and
some  $P_i$'s in the subset $A$, and this codeword is zero at ${\bf
P}-A$ (Lemma 1). Thus we have $|A| \geq n+1-m-1=n-m$. On the other
hand if $A \subset {\bf P}$ and $|A| \geq n-m+2g$ we have that
$dim(L({\bf G}-A^c)\geq deg({\bf G}-A^c)-g+1 \geq g+1$ from
Riemann-Roch theorem (see [17-20]), where $A^c={\bf P}-A$. We also
know that the linear system (see [20]) defined by the divisor ${\bf
G}-A^c$ has no base point since $deg({\bf G}-A^c) \geq 2g$ (see
[20]). Thus we have one function $f \in L({\bf G}-A^c)$ such that
$f$ is not zero at $P_0$ and zero at all points in the set $A^c$, so
the codeword in ${\bf C_L(D,G)}$ corresponding to this $f$ is not
zero at $P_0$ and not
zero at a subset $A$ (or $A$ itself). So $A$ is qualified.\\

{\bf Proof of Proposition 1.} If the LSSS from ${\bf
C_{\Omega}(D,G)}$ is a threshold scheme, it is a $(n-m,m)$ scheme
since $d({\bf C_L(D,G)})=n+1-m$. This imply that any subset $A$ of
${\bf P}$ with cardinality $|A|=m$ is linearly equivalent ($A$ is
considered as a divisor, see [20] for the
 definition of linear equivalence), since $n>3$ and $n-m>1$, we know
 that any two points in ${\bf P}$ are linear equivalent, so
 $dim(L(P_i))
 \geq 2$ for any point $P_i \in {\bf P}$. From Riemann-Roch Theorem
 $dim(L(K-P_i))=dim(L(K))=g$, where $K$ is a canonical divisor of
 the curve ${\bf X}$. If $g=1$, this is obviously not true since
 $K=0$ in this case. If $g \geq 2$, it is known that the canonical
 linear system has no base point. This is a contradiction.\\

 {\bf Proof of Proposition 2}. It is clear $d_{min}$ is the minimum
 distance of the code ${\bf C_L(P,G)}$, so $d_{min} \geq n-m$. On the
 other hand the minimum Hamming weight of ${\bf C_{\Omega}(D,G)}$ is
 at least $m-2g+2$, thus $max _{B \in 2^{\bf P}-\Gamma} |B| \leq m-2g
 $. Thus $d_{cheat}=n-max _{B \in 2^{\bf P}-\Gamma} |B| \leq
 n-m+2g$. The conclusion is proved.\\

{\bf Proof of Theorem 2.} Suppose to secret are distributed, we know
that the shares of the participant $P_i$ are just the function
values $f_1(P_i),f_2(P_i)$, where $f_i$ is a function in $L({\bf
D}-{\bf G}+(\eta))$(corresponding to $f\eta$ in $\Omega({\bf G}-{\bf
D})$). The secrets are just the function values $f_1(P_0),f_2(P_0)$
at $P_0$. Here we have $f_1f_2 \in L(2({\bf D}-{\bf G}+(\eta)))$. If
$2g-2<deg(2( {\bf D}-{\bf G}+(\eta)))<n$, ${\bf C_{\Omega}(D,2(
D-G+(\eta)))}$ is the (non-zero) dual code of ${\bf C_L(D,2( {\bf
D}-{\bf G}+(\eta)))}$. Thus there is a non-zero codeword in ${\bf
C_{\Omega}(D,2( {\bf D}-{\bf G}+(\eta)))}$. On the other hand if the
linear system corresponding to  $\Omega( 2( {\bf D}-{\bf
G}+(\eta)))$ (corresponding to $L(2{\bf G}-{\bf D}-(\eta))$) has no
base point( it is valid if $deg(2{\bf G}-{\bf D}-(\eta)) \geq 2g$,
see [20]), we can make this codeword in ${\bf C_{\Omega}(D,2( {\bf
D}-{\bf G}+(\eta)))}$ not zero at the position $P_0$. Thus
$f_1(P_0)f_2(P_0)$ is a linear combination of the
$f_1(P_1)f_2(P_1),...,f_1(P_n)f_2(P_n)$. The conclusion of 1) is
proved.\\

For the conclusion 2), we only need to prove that the linear system
corresponding to $L(2{\bf G}-{\bf D}-(\eta)-H)$ has no base point
for any $H$ a unqualified subset of ${\bf P}$. From Theorem 1
$deg(H) < n-m+2g$. Thus the conclusion of 2) is true if $deg(2{\bf
G}-{\bf
D}-(\eta)-H) \geq 2g$. The conclusion of 2) is proved.\\

{\bf Proof of Theorem 3.} From Theorem 1 we know that each subset
$H$ of ${\bf P}$ in the adversary structure has at most $n-m+2g-1$
elements. The adversary structure is $Q_2$ if $m \geq
\frac{n}{2}+2g$ and $Q_3$ if $m \geq \frac{2m}{3}+2g$. Then the
conclusions of Theorem 3 follow from Theorem 2 and the main results
in [10] directly.\\

We should note that the parameters $m$ can be chosen quite flexibly
as in the theory of AG-codes( see [17-19]).\\

{\bf III An Asymptotic Result.}\\

For any given finite field $GF(q^2)$ where $q$ is a prime power, it
is known there exists a family of smooth projective curves $\{{\bf
X_t}\}_{(t=1,2,...)}$ defined over $GF(q^2)$ with $N'_t$ rational
points(over $GF(q^2)$) and genus $g_t$ such that $lim
\frac{N'_t}{g_t}=q-1$ (see f.g. [21]), the family of curves over
$GF(q^2)$ attaining the Drinfeld-Vladut bound (see [18-19]). This
family of curves is important for the existence of the family of
AG-codes exceeding the Gilbert-Varshamov bound.  By choosing $m$
suitably we
can have  a similar asymptotic result for the LSSS from AG-codes.\\

{\em {\bf Corollary 1.} For any given finite field $GF(q^2)$ with
$q^2$ ($q \geq 11$ ) elements, there exists a family of natural
numbers $\{N_t\}_{(t=1,2,...)}$ such that $\{N_t\}_{(t=1,2,...)}$ go
to infinity, a family of access structures $\{\Gamma_t\}$ on the set
of $N_t$ elements with the property that any subset less than
$k_t^1$ elements is not in $\Gamma_t$ and any subset more than
$k_t^2$ elements is in $\Gamma_t$ . We can construct\\
1) ideal $GF(q^2)$-LSSS with the access structure $\Gamma_t$;\\
2) VSS over $GF(q^2)$ secure against any adaptive and active
$\Gamma_t^c$ -adversary
structures ($\Gamma_t^c$ consisting of subsets not in $\Gamma_t$);\\
3) MPC (computing any arithmetic circuit over $GF(q^2)$) secure against any adaptive and active $\Gamma_t^c$-adversary.\\
Moreover the parameters $(k_t^1,k_t^2,N_t)$ can be chosen satisfying
$lim \frac{k_t^1}{N_t}=R_1>0$ and $lim\frac{k_t^2}{N_t}=R_2=R_1+\frac{2}{q-1}>0$ for arbitrary
given $R_1 \in (0,\frac{1}{3}-\frac{2}{q-1})$ .}\\

This result follows from the main result in [21] and Theorem 1,2,3
above directly.\\

{\bf IV LSSS from Elliptic Curves}\\

We need to recall the following result in [22-23].\\

{\em {\bf Theorem 4(see [22]).} 1) Let $E$ be an elliptic curve over
$GF(q)$ with the group of $GF(q)$-rational points $E(GF(q))$. Then
$E(GF(q))$ is isomorphic to $Z_{n_1} \bigoplus Z_{n_2}$, where $n_1$
is a divisor of $q-1$ and $n_2$\\
2) If $E$ is supersingular, then $E(GF(q))$ is either\\
a) cyclic; \\
b) or $Z_{2} \bigoplus Z_{\frac{q+1}{2}}$; \\
c) or $Z_{\sqrt q-1} \bigoplus Z_{\sqrt q-1}$;\\
d) or $Z_{\sqrt q+1} \bigoplus Z_{\sqrt q+1}$.}\\

In this section and the section VI we analysis the access structure
of the LSSS from the elliptic curves and the self-dually
$GF(q)$-representable matroid from the
AG-codes on elliptic curves.\\

For any given elliptic curve $E$ over $GF(q)$, from the above
Theorem let $D'=\{0,g_1,...g_{H-1}\}$ be a subgroup of $E(GF(q))$
which is of order $H$(Here $0$ us the zero element of the group).
$0,g_1,...,g_{H-1}$ correspond to the rational points $O
,P_1,P_2,...,P_{H- 1}$ of $E$. In the construction of section II, we
take ${\bf G}=mO$, ${\bf D}=\{P_1,...,P_{H-1}\}$ and ${\bf
P}=\{P_2,...,P_{H-1}\}$. We have the following result.\\

{\em {\bf Theorem 5.} a) Let $A=\{P_{i_1},...,P_{i_m}\}$ be a subset
of ${\bf P}$ with $m$ elements, $A^c$ is a minimal qualified subset
for the LSSS from ${\bf C_{\Omega}(D,G)}$ if and only if
$g_{i_1}+...+g_{i_m} =0$ in $E(GF(q))$;\\
b) Let $A=\{P_{i_1},...,P_{i_{m-1}}\}$ be a subset of ${\bf P}$ with
$m-1$ elements, $A^c$ is a minimal qualified subset for the LSSS
from ${\bf C_{\Omega}(D,G)}$ if and only if
$g_{i_1}+...+g_{i_{m-1}}=0$ in $E(GF(q))$ or there exists a $j \in
\{i_1,...,i_{m-1}\}$ such that
$g_{i_1}+...+g_{i_{m-1}}+g_j=0$ in $E(GF(q))$;\\
c) Any subset of ${\bf P}$ with at least $H-m+2$ elements is
qualified.}\\

{\bf Proof.} We know that for any $t$ points $W_1,...,W_t$ in
$E(GF(q))$ the divisor $W_1+...+W_t- tO$ is linear equivalent to the
divisor $W-O$, where $W$ is the group sum of $W_1,...,W_t$ in the
group $E(GF(q))$. From the proof of Theorem 1,
$\{P_{i_1},...,P_{i_m}\}^c$ is a qualified subset (therefor minimal
qualified subset) if there exist a function $f \in L(G)$ such that
$f(P_{i_1})=...=f(P_{i_m})=0$, this means that the divisor
$P_{i_1}+...+P_{i_m}$ is linearly equivalent to ${\bf G}=mO$.
The conclusion of 1) is proved. \\

From the proof of Theorem 1, $\{P_{i_1},...,P_{i_{m-1}}\}^c$ is a
qualified subset if there exist a function $f \in L(G)$ such that
$f(P_{i_1})=...=f(P_{i_{m-1}})=0$, this means that the divisor
$P_{i_1}+...+P_{i_{m-1}}+B$ is linearly equivalent to ${\bf G}=mO$
for some effective divisor $B$.  It is clear that $deg(B)=1$ and $B$
is a $GF(q)$-rational point in $E$. From the group structure of
$E(GF(q))$, $B$ is in $D'$. On the other hand we note that $B \neq
P_0$, so $B$ is $O$ or a point in ${\bf P}$. The conclusion of 1) is
proved. The conclusion of 3) follows
from Theorem 1 directly.\\

{\bf Example 1. }Let $E$ be the elliptic curve $y^2=x^3+5x+4$
defined over $GF(7)$. Then $E(GF(7))$ is a cyclic group of order
$10$ with $O$ the point at infinity and
$P_0=(3,2),P_1=(2,6),P_2=(4,2),P_3=(0,5)$
$P_4=(5,0),P_5=(0,2),P_6=(4,5),P_7=(2,1),P_8=(3,5)$. From an easy
computation we know that $P_0$ is a generator of $E(GF(7))$ and
$P_i$ is $(i+1)P_0$ (in the group operation of $E(GF(7))$.) We take
${\bf G}=3O,{\bf D}=\{P_0,P_1,...,P_8\}$, then the access structure
of the ideal $GF(7)$-LSSS from ${\bf
C_{\Omega}(D,G)}$ are the following subsets of ${\bf P}=\{P_1,...,P_8\}$.\\
1) All subsets of ${\bf P}$ with $7$ elements and the set ${\bf P}$;\\
2) The following $10$ subsets of $6$ elements $\{P_1,P_3\}^c$
$\{P_1,P_5\}^c$,$\{P_1,P_7\}^c$,$\{P_1,P_8\}^c$
$\{P_2,P_3\}^c$,$\{P_2,P_6\}^c$,\\$\{P_3,P_5\}^c$,$\{P_3,P_7\}^c$,$\{P_5,P_6\}^c$,$\{P_5,P_7\}^c$ are the minimal qualified subset;\\
3) The following $5$ subsets with $5$ elements
$\{P_1,P_2,P_4\}^c$,$\{P_2,P_7,P_8\}^c$,$\{P_3,P_6,P_8\}^c$,
$\{P_4,P_5,P_8\}^c$, $\{P_4,P_6,P_7\}^c$ are the minimal qualified subset.\\

{\bf Example 2.} Let $E$ be the elliptic curve $y^2+y=x^3$ defined
over $GF(4)$. This is the Hermitian curve over $GF(4)$, it has $9$
rational points and $E(GF(4))$ is isomorphic to $Z_3 \bigoplus Z_3$.
We take ${\bf G}=3O$, where $O$ is the zero element in the group
$E(GF(4))$. Let $P_{ij}$ be the rational point on $E$ corresponding
to $(i,j)$ in $Z_3 \bigoplus Z_3$. ${\bf
D}=\{P_{10},P_{01},...,P_{22}\}, {\bf P}=\{P_{01},...,P_{22}\}$.\\
Then the qualified subsets of the ideal LSSS from ${\bf C_{\Omega}(D,G)}$ are as follows.\\
1) The minimal qualified subsets of $4$ elements are
$\{P_{20},P_{21},P_{02}\}^c$,$\{P_{01},P_{20},P_{22}\}^c$,$\{P_{11},P_{12},P_{20}\}^c$.\\
2) The minimal qualified subsets of $5$ elements are
$\{P_{01},P_{02}\}^c$,$\{P_{11},P_{22}\}^c$,$\{P_{12},P_{21}\}^c$.\\
3) The subsets of ${\bf P}$ of $6$ elements and the set ${\bf P}$
are qualified.\\

{\bf V LSSS from Klein Quartic}\\

Klein quartic is the genus $3$ curve $x^3y+y^3z+z^3x=0$ ( in the
projective plane) defined over $GF(8)$. It is well-known there are
$24$ rational points (over $GF(8)$, see [23]).It is clear that
$Q_1=(1:0:0),Q_2=(0:1:0),Q_3=(0:0:1)$ are 3 rational points on ${\bf
X}$. The line $L_0: y=0$ intersects ${\bf X}$ at $3Q_1+Q_3$ (count
with multiplicity, see [20]). The line $L_{\alpha_i}:y=\alpha_i x$,
where $\alpha_1,...,\alpha_7$ are 7 non-zero elements of $GF(8)$,
intersects ${\bf X}$ at $Q_3$ and other $3$ rational points. Set
${\bf P}$ be the set of these $21$ rational points, ${\bf
G}=3Q_1+Q_3$ and ${\bf D}=\{Q_2\} \bigcup {\bf P}$. We consider the
LSSS from the residue code ${\bf C_{\Omega}(D,G)}$.\\

In this case though $deg({\bf G})=2g-2$, ${\bf C_L(D,G)}$ (dimension
3) and ${\bf C_{\Omega}(D,G)}$( dimension 19) are dual codes(see
[19]).\\

It should be noted that the line passing through any two distinct
points in
${\bf P}$ have to pass the other $2$ points in the set ${\bf P}$(see [23]).\\

{\em {\bf Proposition 3.} The minimal qualified subset of the LSSS
from ${\bf C_{\Omega}(D,G)}$ are the subsets of ${\bf P}$ of the
form $\{P_1,P_2,P_3\}^c$, where $P_1,P_2,P_3$ and $Q_3$ are on the
line $L_{\alpha_i}$ for some $i \in \{1,,...,7\}$, or
$\{P'_1,P'_2,P'_3,P'_4\}^c$ for some $4$ points in ${\bf P}$ which
are on one line.}\\

{\bf Proof.} From Riemann-Roch Theorem $dim(L({\bf G}))=3$, so
$\{x,y,z\}$ are the base. So every function in $L({\bf G})$ is of
the form $\frac{ax+by+cz}{y}$. The subset $A^c$ of ${\bf P}$ is
qualified  if and only if there exists one $f \in L({\bf G})$ such
that $f$ is zero on $A$, the conclusion follows directly.\\

{\bf VI Self Dually $GF(q)$ Representable Matroids from AG-codes}\\

For ${\bf X},{\bf D},{\bf G}$ as in section 2, if there exist a
differential ${\eta}$ on ${\bf X}$ such that ${\bf D}-{\bf
G}+(\eta)={\bf G}$ then the matroid from the code $C_L(D,G)$ is
self-dully $GF(q)$-representable. In many cases from algebraic
geometry the matroid of the corresponding AG-codes ${\bf
C_{\Omega}(D,G)}={\bf C_L(D,G)}$ can be calculated.
This offers many examples of new self-dually representable matroids (see [3] for the background).\\

Let $E$ be an elliptic curve defined over $GF(q)$, it is known that
the canonical divisor of $E$ is zero, so the condition that there
exists a differential $\eta$ such that ${\bf D}-{\bf G}+(\eta)={\bf
G}$ is equivalent to the condition that ${\bf D}-{\bf G}$ and ${\bf
G}$ are linear equivalent.\\

Let $D=\{g_1,...g_{H}\}$ be a subset of non-zero elements in
$E(GF(q))$, where $H$ is even and the group sum of all elements in
$D$ is zero in $E(GF(q))$. $g_1,...,g_{H}$ correspond to the
rational points $ P_1,...,P_{H}$ of $E$. In the construction, we
take ${\bf G}=mO$, where $m=\frac{H}{2}$ and $O$ corresponds to the
zero element of $E(GF(q)$, ${\bf D}=\{P_1,...,P_{H}\}$. Here it is
easy to know that $D$ (as a divisor) is linear equivalent to $HO$.
So we know that ${\bf D}-{\bf G}$ is linear equivalent to ${\bf G}$. We have the following result.\\

{\em {\bf Theorem 6.} a) Let $A=\{P_{i_1},...,P_{i_m}\}$ be a subset
of ${\bf D}$ with $m$ elements, $A^c$ is a circuit of the matroid
defined by ${\bf C_{\Omega}(D,G)}={\bf C_L(D,G)}$ if and only if
$g_{i_1}+...+g_{i_m} =0$ in $E(GF(q))$;\\
b) Let $A=\{P_{i_1},...,P_{i_{m-1}}\}$ be a subset of ${\bf D}$ with
$m-1$ elements, $A^c$ is a circuit of the matroid defined by ${\bf
C_{\Omega}(D,G)}={\bf C_L(D,G)}$ if and only if
$g_{i_1}+...+g_{i_{m-1}}=0$ in $E(GF(q))$ or there exists a non-zero
element $g \in (E(GF(q))-{\bf D}) \bigcup A$ such that
$g_{i_1}+...+g_{i_{m-1}}+g=0$ in $E(GF(q))$.}\\

The proof of Theorem 6 is similar to that of Theorem 5.\\

{\bf Example 3. }Let $E$ be the elliptic curve $y^2=x^3+5x+4$
defined over $GF(7)$. Then $E(GF(7))$ is a cyclic group of order
$10$ with $O$ the point at infinity and
$P_0=(3,2),P_1=(2,6),P_2=(4,2),P_3=(0,5)$
$P_4=(5,0),P_5=(0,2),P_6=(4,5),P_7=(2,1),P_8=(3,5)$.Set ${\bf
D}=\{P_0,P_1,P_2,P_3,P_5,P_6,P_7,P_8\}$, ${\bf G}=4O$. It is clear
the group sum of all points in ${\bf D}$ is zero.  Then the circuits
of the self-dually $GF(7)$ representable matroid defined by ${\bf
C_{\Omega}(D,G)}={\bf C_l(D,G)}$ are the following subsets of ${\bf D}$.\\
1) The following $8$ subsets of $4$ elements
$\{P_0,P_1,P_2,P_4\}^c$, $\{P_0,P_1,P_7,P_8\}^c$,
$\{P_0,P_2,P_6,P_8\}^c$,\\ $\{P_0,P_3,P_5,P_8\}^c$,
$\{P_0,P_3,P_6,P_7\}^c$, $\{P_1,P_2,P_6,P_7\}^c$,
$\{P_1,P_3,P_5,P_7\}^c$, $\{P_2,P_3,P_5,P_6\}^c$;\\
3) The following $15$ subsets with $5$ elements
$\{P_0,P_1,P_6\}^c$, $\{P_0,P_2,P_5\}^c$, $\{P_2,P_7,P_8\}^c$,\\
$\{P_3,P_6,P_8\}^c$, $\{P_0,P_5,P_7\}^c$, $\{P_1,P_5,P_6\}^c$,
$\{P_1,P_3,P_8\}^c$, $\{P_2,P_3,P_7\}^c$, $\{P_0,P_1,P_5\}^c$,
$\{P_0,P_5,P_6\}^c$, $\{P_0,P_2,P_7\}^c$, $\{P_1,P_6,P_8\}^c$,
$\{P_1,P_3,P_6\}^c$, $\{P_2,P_5,P_7\}^c$, $\{P_3,P_7,P_8\}^c$.\\

{\bf Example 4.} Let $E$ be the elliptic curve $y^2+y=x^3$ defined
over $GF(4)$. This is the Hermitian curve over $GF(4)$, it has $9$
rational points and $E(GF(4))$ is isomorphic to $Z_3 \bigoplus Z_3$.
We take ${\bf G}=4O$, where $O$ is the zero element in the group
$E(GF(4))$. Let $P_{ij}$ be the rational point on $E$ corresponding
to $(i,j)$ in $Z_3 \bigoplus Z_3$. Let ${\bf G}=4O$ and ${\bf
D}=\{P_{10},P_{01},...,P_{22}\}$ be the $8$ non-zero elements of
$E(GF(4))$. It is clear that ${\bf D}-{\bf G}$ and ${\bf G}$ are
linear equivalent. Then the circuits of the self-dually $GF(4)$
representable  matroid defined by ${\bf
C_{\Omega}(D,G)}={\bf C_L(D,G)}$ are the following subsets of ${\bf D}$.\\
1) The $6$ subsets of $4$ elements are
$\{P_{01},P_{02},P_{10},P_{20}\}^c$,
$\{P_{01},P_{02},P_{11},P_{22}\}^c$,
$\{P_{01},P_{02},P_{12},P_{21}\}^c$,
$\{P_{10},P_{20},P_{11},P_{22}\}^c$,
$\{P_{10},P_{20},P_{12},P_{21}\}^c$, $\{P_{12},P_{21},P_{11},P_{22}\}^c$;\\
2) The $32$ subsets of $5$ elements are
$\{P_{01},P_{10},P_{22}\}^c$, $\{P_{01},P_{11},P_{21}\}^c$,
$\{P_{01},P_{12},P_{20}\}^c$, $\{P_{02},P_{10},P_{21}\}^c$,
$\{P_{02},P_{11},P_{20}\}^c$, $\{P_{02},P_{12},P_{22}\}^c$,
$\{P_{10},P_{11},P_{12}\}^c$, $\{P_{20},P_{21},P_{22}\}^c$,
$\{P_{02},P_{10},P_{22}\}^c$, $\{P_{01},P_{20},P_{22}\}^c$,
$\{P_{01},P_{10},P_{11}\}^c$, $\{P_{02},P_{11},P_{21}\}^c$,
$\{P_{01},P_{22},P_{21}\}^c$, $\{P_{01},P_{11},P_{12}\}^c$,
$\{P_{02},P_{12},P_{20}\}^c$, $\{P_{01},P_{21},P_{20}\}^c$,
$\{P_{01},P_{12},P_{10}\}^c$, $\{P_{01},P_{10},P_{21}\}^c$,
$\{P_{02},P_{20},P_{21}\}^c$, $\{P_{02},P_{10},P_{12}\}^c$,
$\{P_{01},P_{11},P_{20}\}^c$, $\{P_{02},P_{22},P_{20}\}^c$,
$\{P_{02},P_{11},P_{10}\}^c$, $\{P_{01},P_{12},P_{22}\}^c$,
$\{P_{02},P_{12},P_{22}\}^c$, $\{P_{02},P_{11},P_{12}\}^c$,
$\{P_{20},P_{12},P_{11}\}^c$, $\{P_{10},P_{22},P_{12}\}^c$,
$\{P_{10},P_{11},P_{21}\}^c$, $\{P_{10},P_{21},P_{22}\}^c$,
$\{P_{20},P_{12},P_{22}\}^c$, $\{P_{20},P_{21},P_{11}\}^c$.\\

{\bf VII Conclusions} \\

We have presented the ideal linear secret-sharing scheme from the
AG-codes on algebraic curves, which can be thought as a natural
generalization of Shamir's $(k,n)$-threshold scheme(from AG-codes on
the genus 0 curve, RS codes). These ideal linear secret-sharing
schemes are not threshold for positive genus curves, which offer
many
 new examples of access structures of ideal LSSS. The general properties
  of LSSS from AG-codes are proved and their applications in
  verifiable secret-sharing and secure multi-party computation are presented.
 New examples of self-dually representable matroids from self-dual AG-codes
 have been calculated. We demonstrated that the algebraic-geometric structure
 of the underlying curves is an important resource for secret-sharing,
  multi-party computation and the theory of matroids.    \\

{\bf Note.} {\em After this paper was completed and submitted, the
author was informed by Professor R. Cramer of his paper"Algebraic
geometric secret sharing schemes and secure computation over small
fields", in which the idea of using algebraic-geometric codes in
secret sharing and
secure computation was independently developed.}\\

{\bf Acknowledgement.} This work is supported by Grant 60542006 and
Distinguished Young Scholar Grant 10225106 of NNSF, China.\\

\begin{center}
REFERENCES
\end{center}

[1] E.F. Brickell,  Some ideal secret sharing schemes, in Advances
in Cryptology- Eurocrypt'89, Lecture notes in Computer Science, vol.
434, Springer, Heidelberg, 1990, pp.468-475\\

[2] R.J.McEliece and D.V.Sarwate, On sharing secrets and Reed-Solomom codes, Comm. ACM, 22,11(Nov.1979), pp.612-613\\

[3] R. Cramer, V.Daza,I.Cracia, J.J. Urroz, C.Leander, J.Marti-Farre
and C.Padro, On Codes, matroids and secure multi-party computations
from linear secret sharing schems, Advances in Cryptology, Crpto
2005, LNCS 3621, pp327-343.\\

[4] G.R.Blakle, safeguarding cryptographic keys, Proc. NCC
AFIPS(1979), pp.313-317\\

[5] A.Shamir, How to share a secret, Comm. ACM 22(1979), pp.612-613\\

[6] J.Benaloh and J.Leichter, Generalized secret sharing and
monotone functions, Crypto'88, LNCS-403, pp.25-35\\

[7] A.Ito, A.Saito and T.Nishizeki, Secret sharing scheme realizing
general access structures, Proc. IEEE Globalcom'1987, Tokyo,
pp.99-102\\

[8] K.Okada and K.Kurosawa, MDS secret-sharing scheme secure against
cheaters, IEEE Transactions on Information Theory,
Vol.46(2000), no.3, pp.1078-81\\

[9] J.D.Golic, On matroid characterization of ideal secret sharing
schemes, J. Cryptology, Vol.11(1998), pp.75-86\\

[10] R. Cramer, I.Damgard and U.Maurer, General secure multi-party
computation from any linear secret-sharing scheme, Eurocrypto 2000,
LNCS-1807, pp.316-334\\

[11] R.Cramer, Introduction to secure computation,\\
http://www.inf.ethz.ch/personal/cramer\\

[12] J.L.Massey, Minimal codewords and secret sharing, Proc. 6th
Joint Sweidish-Russsian workshop on Information Theory, Molle,
Sweden,August 22-27, 1993,pp269-279 \\

[13] J.L.Massey, Some applications of coding theory in cryptography,
in P.G.Farrell (Ed.0, Codes and Ciphers: Cryptography and Coding IV,
Formara Ltd, Essses, England, 1995, pp.33-47\\

[14] C.Ding, D.R.Kohel and S.Ling, Secret-sharing with a class of
tenary codes, Theoretical Computer Science, Vol.246(2000),
pp.285-298\\

[15] R.J.Anderson, C.Ding, T.Helleseth and T.Klove, How to build
robust shared control system, Designs,
 Codes and Cryptography, Vol. 15(1998), pp.111-124  \\

[16] J.H. van Lint and R.M.Wilson, A course in combinatorics,
Cambridge Univ. Press, 2001\\

[17] J.H.van Lint, Introduction to coding theory (3rd Edition),
Springer-Verlag, 1999\\

[18] M.A.Tsfasman and S.G.Vladut, Algebraic-geometric codes, Kluwer,
Dordrecht, 1991\\

[19] H.Stichtenoth, Algebraic function fields and codes, Springer,
Berlin, 1993\\

[20] R.Hartshorne, Algebraic geometryGTM 52, Springer-Verlag, 1977\\

[21] A.Gacia and H.Stichtenoth, A tower of Artin-Schreier extension
of function fields attaining Drinfeld-Vladut bound, Invent. Math.,
121(1995), no.1, pp211-222\\

[22] R.Schoof, Nonsingular plane cubic curves over finite fields,
J.Combin. Theory, A, vol.46(1987). pp.183-211\\

[23] D.Hankerson, A.Menezes and S.Vanstone, Guide to elliptic curve
cryptography, Springer-Verlag, 2004\\

\end{document}